\documentclass[prd,aps,nofootinbib,floatfix,11pt]{revtex4}

\usepackage{amsmath,graphicx,epsfig,amssymb,dsfont,mathtools}
\usepackage[usenames]{color}
\usepackage{ulem} 
\usepackage{bigstrut}
\usepackage{slashed}
\usepackage{multirow}
\usepackage{subfigure}

\allowdisplaybreaks

\begin{document}
	
\title{The SVZ sum rules and the heavy quark limit for $\Lambda_{Q}$}
	
\author{
Zhen-Xing Zhao$^{1}$~\footnote{Email:zhaozx19@imu.edu.cn}, 
Run-Hui Li$^{1}$~\footnote{Email:lirh@imu.edu.cn}, 
Yu-Ji Shi$^{2}$~\footnote{Email:shiyuji92@126.com}, 
Si-Hong Zhou$^{1}$
}
	
\affiliation{
$^{1}$ School of Physical Science and Technology, \\
Inner Mongolia University, Hohhot 010021, China \\
$^{2}$ Helmholtz-Institut f\"ur Strahlen- und Kernphysik and Bethe Center
for Theoretical Physics, Universit\"at Bonn, 53115 Bonn, Germany
}
		
\begin{abstract}
In this work, we evaluate the accuracy of the leading order results in Shifman-Vainshtein-Zakharov (SVZ) sum rules and the leading power results in the heavy quark limit for the mass of $\Lambda_{Q}$. Up to dim-5 condensate contributions are considered. By comparing with the experimental results, we find that the leading order results in SVZ sum rules can reach about 5\% accuracy both for $\Lambda_{b}$ and $\Lambda_{c}$, and it seems that better results can be obtained from the sum rules of the heavy quark limit both for $\Lambda_{b}$ and $\Lambda_{c}$. We re-check that the SVZ sum rules in the heavy quark limit coincide with the HQET sum rules. As a byproduct, we also have an exploratory discussion on the $ud$ diquark in $\Lambda_{Q}$.
\end{abstract}

\maketitle
	
\section{Introduction}

In 2018, the LHCb Collaboration updated the measurement of the lifetime
of the charmed baryon $\Omega_{c}^{0}$ \cite{Aaij:2018dso}, which is nearly 4 times larger
than the world average in PDG2018 \cite{Tanabashi:2018oca}. On the theoretical side,
some attempts have been made to solve this puzzle \cite{Cheng:2018rkz}.
However, there is still a lack of QCD based calculation. Using the
SVZ sum rules and following the line sketched in our previous
work \cite{Shi:2019hbf}, we preliminarily obtain the matrix elements of the
four-quark operators which arise from the heavy quark expansion of
hadron lifetime. The study on the lifetimes of heavy baryons will
appear in our next few articles.

In most cases, one can only obtain the leading order expressions in
the SVZ sum rules, and this situation is expected to last for a long
time in the light of the great challenge from radiation correction
calculation. The leading order results are sensitive to the heavy quark 
mass, thus, how to choose the heavy quark mass becomes a problem.
In the history of the application of the SVZ sum rules, there are
two different viewpoints on the selection of heavy quark mass. One choice
is the pole mass, and the other is the current mass. These two schemes may lead to very different results. For the calculation of the decay constant of $B$ meson, as pointed out by Ref. \cite{Jamin:2001fw}, in the pole mass scheme the first two order corrections do not show any sign of convergence.
In this work, we will have to be content with the leading order calculation,
and take the $\overline{{\rm MS}}$ mass for heavy quark. 
One way to avoid choosing the heavy quark mass is to
take the heavy quark limit, under which, the results will not depend
on the heavy quark mass. From Refs. \cite{Shuryak:1981fza, Colangelo:1995qp},
we can see that the heavy quark limit can lead to the HQET
sum rules for the case of singly heavy hadron. At this point, there
are still two questions: 1) How good is the leading order approximation
in the SVZ sum rules? 2) How good is the leading power approximation
in the heavy quark limit?

The first question is partly answered by Ref. \cite{Jamin:2001fw}, in which
the LO result contributes the most and the NLO and NNLO results show good
convergence for decay constant of $B$ meson
$f_{B}$. The second question is partly answered by Ref. \cite{Colangelo:1995qp},
in which, the binding energy of light quarks was obtained for $\Lambda_{Q}$.
However, this result was not applied to the charmed baryon $\Lambda_{c}$, to which,
question 2) above is expected to be very important. In this work, we will investigate the masses of heavy baryons
at the leading order and at the leading power, and the corresponding results are then compared with the
experimental values to evaluate how good these approximations are.

The rest of the paper is arranged as follows. In Sect.~II, the SVZ sum rules and the heavy quark limit are shown, including the definition of the interpolating current for $\Lambda_{Q}$ and the results for the QCD spectral density. Numerical results are given in Sect. III. A exploratory discussion on the $ud$ diquark in $\Lambda_{Q}$ can be found in Sect. IV. We conclude this paper in the last section.

\section{The SVZ sum rules and the heavy quark limit}

Similar as that in Ref. \cite{Wang:2010fq}, we will adopt the simplest
interpolating current for $\Lambda_{Q}$
\begin{equation}
J=\frac{1}{\sqrt{2}}\epsilon_{abc}(u_{a}^{T}C\gamma_{5}d_{b}-d_{a}^{T}C\gamma_{5}u_{b})Q_{c}.\label{eq:interpolating_current}
\end{equation}
More reliable results can be obtained by considering other possible
interpolating currents for the baryon, however, it is not the purpose
of this work.

The definition of the two-point correlation function is standard
\begin{equation}
\Pi(p)=i\int d^{4}x\ e^{ip\cdot x}\langle0|T\{J(x)\bar{J}(0)\}|0\rangle.
\end{equation}
On the hadronic side, one can insert the complete set of hadronic
states, and the correlation function is then written as 
\begin{equation}
\Pi^{{\rm had}}(p)=\lambda_{+}^{2}\frac{\slashed p+M_{+}}{M_{+}^{2}-p^{2}}+\lambda_{-}^{2}\frac{\slashed p-M_{-}}{M_{-}^{2}-p^{2}}+\cdots,
\end{equation}
where we have also considered the contribution from the negative parity
baryon, $M_{\pm}$ ($\lambda_{\pm}$) stand for the masses (the pole
residues) of the positive and negative parity baryons. It turns out
that it is trivial to consider the contribution from the negative
parity baryon for the present case, but it is not for the doubly heavy
baryon case \cite{Hu:2017dzi}. The negative-parity contribution was discussed in \cite{Khodjamirian:2011jp}.

On the QCD side, the correlation function can be calculated following
standard procedures, and the result can be formally written as 
\begin{equation}
\Pi^{{\rm QCD}}(p)=A(p^{2})\slashed p+B(p^{2}).
\end{equation}
The coefficients $A$ and $B$ can be further written into the form
of dispersion relation
\begin{equation}
A(p^{2})=\int ds\ \frac{\rho^{A}(s)}{s-p^{2}},\quad B(p^{2})=\int ds\ \frac{\rho^{B}(s)}{s-p^{2}}.\label{eq:rho_A_rho_B}
\end{equation}
The specific expressions of the spectral density $\rho^{A}$ and $\rho^{B}$
have been obtained in Ref. \cite{Wang:2010fq}. For convenience, we
quote these results as follows\footnote{Here we have neglected the contribution from $\langle\bar{q}q\rangle^{2}$
and a factor of 2 has been considered since our interpolating current
for the baryon is $\sqrt{2}$ times of that in Ref. \cite{Wang:2010fq}.}
\begin{eqnarray}
\rho^{A}(p^{2}) & = & \frac{3}{64\pi^{4}}\int_{t_{i}}^{1}dt\ t(1-t)^{2}(p^{2}-\tilde{m}_{Q}^{2})^{2}\nonumber \\
 &  & +\frac{1}{64\pi^{2}}\langle\frac{\alpha_{S}}{\pi}GG\rangle\int_{t_{i}}^{1}dt\ t\nonumber \\
 &  & -\frac{1}{192\pi^{2}}\langle\frac{\alpha_{S}}{\pi}GG\rangle(1-t_{i})^{2},\label{eq:rho_A}
\end{eqnarray}
\begin{eqnarray}
\rho^{B}(p^{2}) & = & \frac{3m_{Q}}{64\pi^{4}}\int_{t_{i}}^{1}dt\ (1-t)^{2}(p^{2}-\tilde{m}_{Q}^{2})^{2}\nonumber \\
 &  & +\frac{m_{Q}}{64\pi^{2}}\langle\frac{\alpha_{S}}{\pi}GG\rangle\int_{t_{i}}^{1}dt\nonumber \\
 &  & +\frac{m_{Q}}{96\pi^{2}}\langle\frac{\alpha_{S}}{\pi}GG\rangle\int_{t_{i}}^{1}dt\ \frac{(1-t)^{3}}{t^{2}}\nonumber \\
 &  & -\frac{m_{Q}}{192\pi^{2}}\langle\frac{\alpha_{S}}{\pi}GG\rangle(1-t_{i})^{2}.\label{eq:rho_B}
\end{eqnarray}
Here $\tilde{m}_{Q}^{2}\equiv m_{Q}^{2}/t$, $t_{i}\equiv m_{Q}^{2}/p^{2}$,
and the contributions from dim-3 quark condensate and dim-5 quark-glon
mixed condensate vanish. 

In the following, we will consider the heavy quark limit of $\rho^{A}$
and $\rho^{B}$. In the heavy quark limit, all the terms of $\rho^{A}$
when multiplied by the mass of $\Lambda_{Q}$ respectively scale as
\begin{equation}
\frac{\omega^{5}}{20\pi^{4}},\quad\frac{\omega}{32\pi^{2}}\langle\frac{\alpha_{S}}{\pi}GG\rangle,\quad-\frac{\omega^{2}}{48\pi^{2}m_{Q}}\langle\frac{\alpha_{S}}{\pi}GG\rangle,
\end{equation}
where $m_{\Lambda_{Q}}=m_{Q}+\Delta$ with $\Delta$ the binding energy
of light quarks has been used. All the terms of $\rho^{B}$ respectively
scale as
\begin{equation}
\frac{\omega^{5}}{20\pi^{4}},\quad\frac{\omega}{32\pi^{2}}\langle\frac{\alpha_{S}}{\pi}GG\rangle,\quad\frac{\omega^{4}}{24\pi^{2}m_{Q}^{3}}\langle\frac{\alpha_{S}}{\pi}GG\rangle,\quad-\frac{\omega^{2}}{48\pi^{2}m_{Q}}\langle\frac{\alpha_{S}}{\pi}GG\rangle.
\end{equation}
It can be seen that, there are many different scales for the gluon
condensate contributions. In the heavy quark limit, $m_{\Lambda_{Q}}\rho^{A}$
and $\rho^{B}$ are all reduced to
\begin{equation}
\rho(\omega)\equiv\frac{\omega^{5}}{20\pi^{4}}+\frac{\omega}{32\pi^{2}}\langle\frac{\alpha_{S}}{\pi}GG\rangle.
\end{equation}

The SVZ sum rules are given by 
\begin{align}
(M_{+}+M_{-})\lambda_{+}^{2}\exp(-M_{+}^{2}/T_{+}^{2}) & =\int_{m_{Q}^{2}}^{s_{+}}ds\ (M_{-}\rho^{A}+\rho^{B})\exp(-s/T_{+}^{2}),\nonumber \\
(M_{+}+M_{-})\lambda_{-}^{2}\exp(-M_{-}^{2}/T_{-}^{2}) & =\int_{m_{Q}^{2}}^{s_{-}}ds\ (M_{+}\rho^{A}-\rho^{B})\exp(-s/T_{-}^{2}),\label{eq:SVZ_sum_rules}
\end{align}
where $T_{\pm}^{2}$ are the Borel parameters and $s_{\pm}$ are the
continuum threshold parameters. In the heavy quark limit, the sum
rule is given by 
\begin{equation}
\frac{1}{2}\lambda_{+}^{2}\exp(-\Delta_{+}/E_{+})=\int_{0}^{\omega_{+}}d\omega\ \rho(\omega)\exp(-\omega/E_{+}),\label{eq:heavy_quark_limit_sum_rule}
\end{equation}
where $E_{+}$ is the Borel parameter and $\omega_{+}$ is the continuum
threshold parameter. Note that the counterpart for $\lambda_{-}$
vanishes in the heavy quark limit for the interpolating current defined
in Eq. (\ref{eq:interpolating_current}).

Respectively differentiating the first equation of Eqs. (\ref{eq:SVZ_sum_rules})
with $-1/T_{+}^{2}$ and Eq. (\ref{eq:heavy_quark_limit_sum_rule})
with $-1/E_{+}$, one can arrive at the sum rule 
\begin{equation}
M_{+}^{2}=\frac{\int_{m_{Q}^{2}}^{s_{+}}ds\ (M_{-}\rho^{A}+\rho^{B})\ s\ \exp(-s/T_{+}^{2})}{\int_{m_{Q}^{2}}^{s_{+}}ds\ (M_{-}\rho^{A}+\rho^{B})\ \exp(-s/T_{+}^{2})}
\end{equation}
for the mass and the sum rule 
\begin{equation}
\Delta_{+}=\frac{\int_{0}^{\omega_{+}}d\omega\ \rho(\omega)\ \omega\ \exp(-\omega/E_{+})}{\int_{0}^{\omega_{+}}d\omega\ \rho(\omega)\ \exp(-\omega/E_{+})}
\end{equation}
for the binding energy. Numerical results will be
given in the next section.

\section{Numerical results}

We do not intend to give an elaborative determination of baryon mass, but
merely give an estimate of the result from the leading order or the
leading power. For our purpose, in the numerical analysis, we will
only consider the centural values for the inputs \cite{Tanabashi:2018oca,Colangelo:2000dp}:
$m_{b}=4.18\ {\rm GeV}$, $m_{c}=1.27\ {\rm GeV}$, $\langle\frac{\alpha_{S}}{\pi}GG\rangle=0.012\ {\rm GeV}^{4}$.

\begin{figure}[!]
\includegraphics[width=1.0\columnwidth]{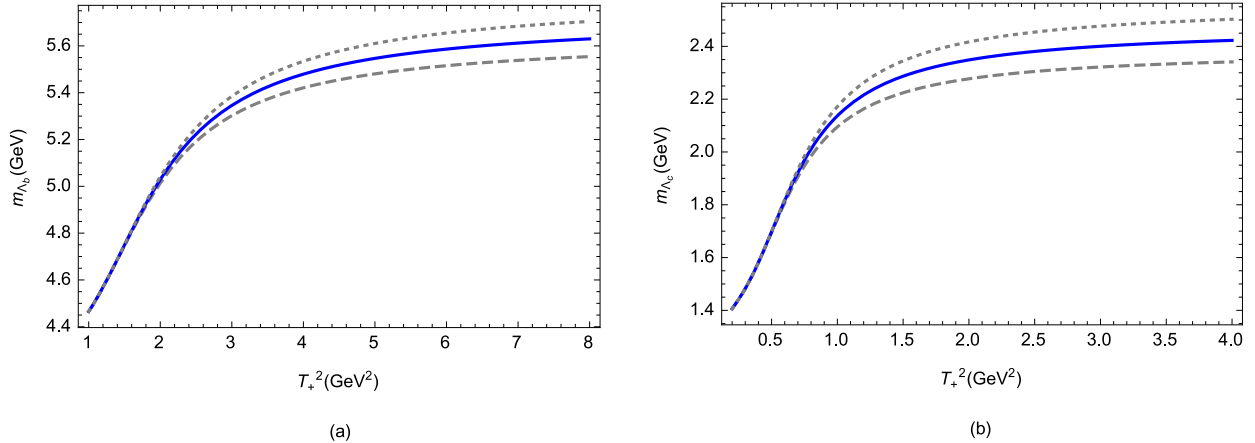}
\caption{The masses for $\Lambda_{b}$ and $\Lambda_{c}$ as a function of the Borel parameter $T_{+}^{2}$. The dashed line, the solid line, and the dotted line respectively correspond to $s_{+} = 5.9^{2}, 6.0^{2}, 6.1^{2} \ {\rm GeV}^{2}$ for $\Lambda_{b}$ and $s_{+} = 2.6^{2}, 2.7^{2}, 2.8^{2} \ {\rm GeV}^{2}$ for $\Lambda_{c}$.}
\label{fig:SVZ} 
\end{figure}

In the SVZ sum rules, the results for $m_{\Lambda_{b}}$ and $m_{\Lambda_{c}}$
as a function of the Borel parameter $T_{+}^{2}$ are shown in Fig. \ref{fig:SVZ}, where the continuum threshold parameter $\sqrt{s_{+}}$ is chosen as $0.3-0.5\ {\rm GeV}$ larger than the mass of the ground state.
For $\Lambda_{b}$, the Borel window can be chosen as $3-4\ {\rm GeV}^{2}$, and then the mass of $\Lambda_{b}$ is determined as 
\begin{equation}
m_{\Lambda_{b}}=5.4\pm0.1\ {\rm GeV}.
\end{equation}
Compared with the experimental value $m_{\Lambda_{b}}=5.620\ {\rm GeV}$
\cite{Tanabashi:2018oca}, we can see a deviation of about 5\%, which would be
expected, because here we have only considered the leading order calculation
of the QCD spectral desity. In Ref. \cite{Jamin:2001fw} which considered
the decay constant of $B$ meson, this deviation is about
10\%. For $\Lambda_{c}$, the Borel window can be chosen as $1.0-1.5\ {\rm GeV}^{2}$,
and then the mass of $\Lambda_{c}$ is determined as 
\begin{equation}
m_{\Lambda_{c}}=2.2\pm0.1\ {\rm GeV}.
\end{equation}
Compared with the experimental value $m_{\Lambda_{c}}=2.286\ {\rm GeV}$ \cite{Tanabashi:2018oca}, we can see a deviation of about 5\%.

\begin{figure}[!]
\includegraphics[width=0.5\columnwidth]{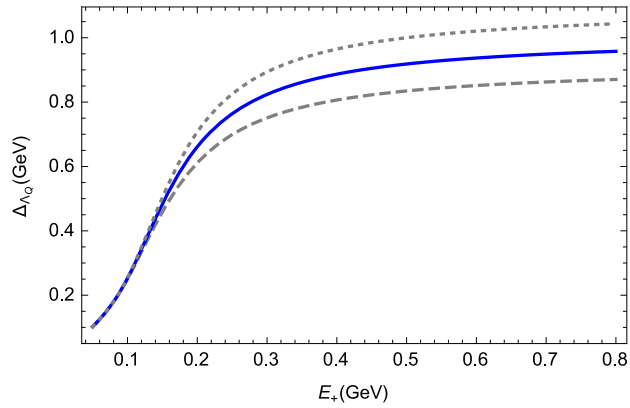}
\caption{The binding energy of $\Lambda_{Q}$ as a function of the Borel parameter $E_{+}$. The dashed line, the solid line, and the dotted line respectively correspond to $\omega_{+} = 1.1, 1.2, 1.3 \ {\rm GeV}$.}
\label{fig:HQL} 
\end{figure}

In the sum rule of the heavy quark limit, the result for the binding
energy of $\Lambda_{Q}$ as a function of the Borel parameter $E_{+}$
is shown in Fig. \ref{fig:HQL}, where the continuum threshold parameter $\omega_{+}$
is chosen as $1.1-1.3\ {\rm GeV}$. The Borel window can be chosen
as $0.3-0.4\ {\rm GeV}$, and then the binding energy of $\Lambda_{Q}$
is determined as 
\begin{equation}
\Delta_{\Lambda_{Q}}=0.9\pm0.1\ {\rm GeV}.
\end{equation}
We can further determine the masses of $\Lambda_{b}$ and $\Lambda_{c}$ as 
\begin{align}
m_{\Lambda_{b}} & =m_{B}-\Delta_{B}+\Delta_{\Lambda_{Q}}\approx5.7\ {\rm GeV},\nonumber \\
m_{\Lambda_{c}} & =m_{D}-\Delta_{D}+\Delta_{\Lambda_{Q}}\approx2.3\ {\rm GeV},
\end{align}
where the binding energy for $B$ meson $\Delta_{B}\approx0.5\ {\rm GeV}$ \cite{Ball:1993xv} and $\Delta_{D}\approx\Delta_{B}$ in the heavy quark limit have been used.

\section{An exploratory discussion on the diquark in $\Lambda_{Q}$}

\begin{figure}[!]
\includegraphics[width=0.5\columnwidth]{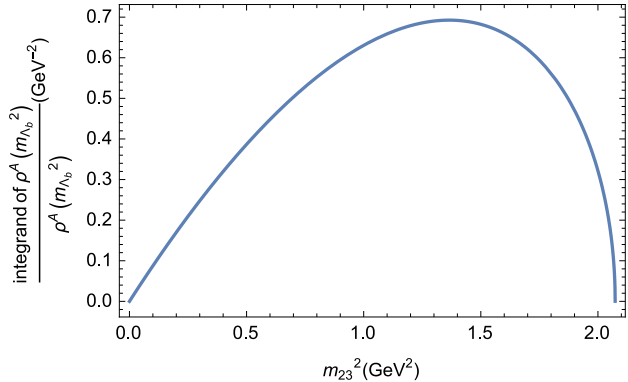}
\caption{The invariant mass spectrum of the $ud$ diquark in $\Lambda_{b}$. Here $\rho^{A}$ in Eq. (\ref{eq:rho_A_rho_B_cutting_rule}) is considered.}
\label{fig:m23s} 
\end{figure}

The discussion in this section is exploratory in nature. When constructing
the interpolating current in Eq. (\ref{eq:interpolating_current})
for $\Lambda_{Q}$, one has considered the $u$ and $d$ quarks as
a scalar diquark system. Using the cutting rule, one can obtain the
QCD spectral densities $\rho^{A}$ and $\rho^{B}$ in Eq. (\ref{eq:rho_A_rho_B})
as:
\begin{align}
\rho^{A}(s) & =\frac{3}{8\pi^{6}}\int_{0}^{(\sqrt{s}-m_{1})^{2}}dm_{23}^{2}\ m_{23}^{2}\ \frac{s+m_{1}^{2}-m_{23}^{2}}{2s}\ \frac{\pi\sqrt{\lambda(s,m_{1}^{2},m_{23}^{2})}}{2s}\frac{\pi}{2},\nonumber \\
\rho^{B}(s) & =\frac{3}{8\pi^{6}}\int_{0}^{(\sqrt{s}-m_{1})^{2}}dm_{23}^{2}\ m_{23}^{2}\ m_{1}\ \frac{\pi\sqrt{\lambda(s,m_{1}^{2},m_{23}^{2})}}{2s}\frac{\pi}{2}.\label{eq:rho_A_rho_B_cutting_rule}
\end{align}
Here we only consider the perturbative contribution, $m_{1}\equiv m_{Q}$
is the heavy quark mass, $m_{23}^{2}$ is the invariant mass squared
of the $ud$ diquark system, and $\lambda(a,b,c)\equiv a^{2}+b^{2}+c^{2}-2(ab+bc+ca)$. In the rest of this section, we will only consider the spectral density $\rho^{A}$ in Eq. (\ref{eq:rho_A_rho_B_cutting_rule}), and similar results can be obtained for $\rho^{B}$. The counterpart of $\rho^{A}$ at the hadronic side is proportional to $\lambda_{+}^{2}\delta(s-M_{+}^{2})$, where we have only considered the contribution from the positive parity baryon. However, one can see more information from the QCD spectral density $\rho^{A}$: the invariant mass squared of the $ud$ diquark shows some kind of distribution. Fixing $s=m_{\Lambda_{b}}^{2}$, the integrand of $\rho^{A}$ is plotted in Fig. \ref{fig:m23s}. 

It can be seen from Fig. \ref{fig:m23s} that, the distribution of $m_{23}^{2}$
ranges from $0$ to $(m_{\Lambda_{b}}-m_{b})^{2}$, and peaks at about
$m_{23}^{2}=(1.17\ {\rm GeV})^{2}$. Some comments are in order.
\begin{itemize}
\item The two zeros of the integrand of $\rho^{A}$ in Eq. (\ref{eq:rho_A_rho_B_cutting_rule})
respectively come from the $m_{23}^{2}$ term and the two-body phase
space of the heavy quark and the diquark. Furthermore, the $m_{23}^{2}$
term comes from the inner product of the 4-momenta of $u$ quark and $d$ quark. 
It would be expected to contain some information of the diquark.
\item In this work, we have neglected the masses of $u$ and $d$ quarks.
The most probable mass indicates that the $ud$ diquark is energetic in the baryon.
\end{itemize}

\section{Conclusions}

In 2018, the LHCb updated the measurement for the lifetime of $\Omega_{c}^{0}$,
great deviation has been found compared with the world average in
PDG2018. A solid theoretical calculation is highly demanded to clarify
this matter.

The lifetimes of heavy hadrons are usually analyzed using the heavy
quark expansion (HQE). In HQE, the total decay width of heavy hadron
is expanded in powers of $1/m_{Q}$ :
\begin{equation}
\Gamma(H_{Q}\to f)=\frac{G_{F}^{2}m_{Q}^{5}}{192\pi^{3}}V_{{\rm CKM}}\left(A_{0}+\frac{A_{2}}{m_{Q}^{2}}+\frac{A_{3}}{m_{Q}^{3}}+\frac{A_{4}}{m_{Q}^{4}}+{\cal O}\left(\frac{1}{m_{Q}^{5}}\right)\right),\label{eq:HQE}
\end{equation}
where $V_{\rm CKM}$ is the relevant CKM matrix element. As pointed out in Ref. \cite{Cheng:2018rkz}, at present, the major uncertainties
come from the matrix elements of the four-quark operators in $1/m_{Q}^{3}$ term. We have obtained some preliminary results for these matrix elements using QCD sum rules. However,
there are still some questions we have to answer. Since we can only
perform the LO calculation, it is necessary to evaluate how good the
LO approximation is in the SVZ sum rules. It would also be helpful
to explore the heavy quark limit, from which one can gain insight
into the contribution of each condensate. Naturally, we need to evaluate the accuracy of the leading power in the heavy quark limit. By taking the heavy quark limit of the SVZ sum rules, we arrive at the HQET sum rules for the singly heavy baryon case. It would be especially valuable to explore this limit for the doubly heavy baryon case, since in this situation, HQET is not applicable.

Through this study, we find that the LO calculation in the SVZ sum rules
can reach about 5\% accuracy for the masses of $\Lambda_{b}$ and
$\Lambda_{c}$, and it seems that better results can be obtained
using the sum rules in the heavy quark limit. For
the matrix elements of the four-quark operators in $1/m_{Q}^{3}$
term of Eq. (\ref{eq:HQE}), it is enough if the LO calculation can reach similar accuracy. As a byproduct, we also have an exploratory discussion on the $ud$ diquark in $\Lambda_{Q}$.

In 2017, the LHCb Collaboration reported the discovery of doubly charmed
baryon $\Xi_{cc}^{++}$, which has triggered a wide interest in studying
the properties of doubly heavy baryons \cite{Wang:2017mqp,Meng:2017udf,Wang:2017azm,
Gutsche:2017hux,Li:2017pxa,Guo:2017vcf,Lu:2017meb,Xiao:2017udy,Sharma:2017txj,Ma:2017nik,Yu:2017zst,Meng:2017dni,Hu:2017dzi,Cui:2017udv,Shi:2017dto,Xiao:2017dly,Yao:2018zze,Yao:2018ifh,
Ozdem:2018uue,Ali:2018ifm,Dias:2018qhp,Li:2018epz,Zhao:2018mrg,Xing:2018bqt,Zhu:2018epc,
Ali:2018xfq,Liu:2018euh,Xing:2018lre,Bediaga:2018lhg,Wang:2018duy,Dhir:2018twm,Berezhnoy:2018bde,Jiang:2018oak,Zhang:2018llc,Li:2018bkh,Meng:2018zbl,Cerri:2018ypt,Gutsche:2018msz,Shi:2019hbf,Shi:2019fph,Shi:2020qde,Hu:2020mxk}. The
SVZ sum rules and/or the heavy quark limit can be performed on the masses,
form factors and lifetimes to get a better understanding of double heavy baryons.

\section*{Acknowledgements}

The authors are grateful to Profs.~Ying Li, Wei Wang, Yu-Ming Wang, Zhi-Gang Wang and Fu-Sheng Yu for valuable comments and discussions. This work is supported in part by National Natural Science Foundation of China under Grants No. 11765012, 11947414.


\begin{thebibliography}{1}

\bibitem{Aaij:2018dso} 
  R.~Aaij {\it et al.} [LHCb Collaboration],
  Phys.\ Rev.\ Lett.\  {\bf 121}, no. 9, 092003 (2018)
  doi:10.1103/PhysRevLett.121.092003
  [arXiv:1807.02024 [hep-ex]].


\bibitem{Tanabashi:2018oca} 
  M.~Tanabashi {\it et al.} [Particle Data Group],
  Phys.\ Rev.\ D {\bf 98}, no. 3, 030001 (2018).
  doi:10.1103/PhysRevD.98.030001


\bibitem{Cheng:2018rkz} 
  H.~Y.~Cheng,
  JHEP {\bf 1811}, 014 (2018)
  doi:10.1007/JHEP11(2018)014
  [arXiv:1807.00916 [hep-ph]].


\bibitem{Shi:2019hbf} 
  Y.~J.~Shi, W.~Wang and Z.~X.~Zhao,
  arXiv:1902.01092 [hep-ph].


\bibitem{Jamin:2001fw} 
  M.~Jamin and B.~O.~Lange,
  Phys.\ Rev.\ D {\bf 65}, 056005 (2002)
  doi:10.1103/PhysRevD.65.056005
  [hep-ph/0108135].


\bibitem{Shuryak:1981fza} 
  E.~V.~Shuryak,
  Nucl.\ Phys.\ B {\bf 198}, 83 (1982).
  doi:10.1016/0550-3213(82)90546-6


\bibitem{Colangelo:1995qp} 
  P.~Colangelo, C.~A.~Dominguez, G.~Nardulli and N.~Paver,
  Phys.\ Rev.\ D {\bf 54}, 4622 (1996)
  doi:10.1103/PhysRevD.54.4622
  [hep-ph/9512334].


\bibitem{Wang:2010fq} 
  Z.~G.~Wang,
  Eur.\ Phys.\ J.\ C {\bf 68}, 479 (2010)
  doi:10.1140/epjc/s10052-010-1365-8
  [arXiv:1001.1652 [hep-ph]].


\bibitem{Hu:2017dzi} 
  X.~H.~Hu, Y.~L.~Shen, W.~Wang and Z.~X.~Zhao,
  Chin.\ Phys.\ C {\bf 42}, no. 12, 123102 (2018)
  doi:10.1088/1674-1137/42/12/123102
  [arXiv:1711.10289 [hep-ph]].


\bibitem{Khodjamirian:2011jp} 
  A.~Khodjamirian, C.~Klein, T.~Mannel and Y.-M.~Wang,
  JHEP {\bf 1109}, 106 (2011)
  doi:10.1007/JHEP09(2011)106
  [arXiv:1108.2971 [hep-ph]].


\bibitem{Colangelo:2000dp} 
  P.~Colangelo and A.~Khodjamirian,
  In *Shifman, M. (ed.): At the frontier of particle physics, vol. 3* 1495-1576
  doi:10.1142/9789812810458\_0033
  [hep-ph/0010175].


\bibitem{Ball:1993xv} 
  P.~Ball and V.~M.~Braun,
  Phys.\ Rev.\ D {\bf 49}, 2472 (1994)
  doi:10.1103/PhysRevD.49.2472
  [hep-ph/9307291].


\bibitem{Wang:2017mqp} 
  W.~Wang, F.~S.~Yu and Z.~X.~Zhao,
  Eur.\ Phys.\ J.\ C {\bf 77}, no. 11, 781 (2017)
  doi:10.1140/epjc/s10052-017-5360-1
  [arXiv:1707.02834 [hep-ph]].


\bibitem{Meng:2017udf} 
  L.~Meng, N.~Li and S.~l.~Zhu,
  Eur.\ Phys.\ J.\ A {\bf 54}, no. 9, 143 (2018)
  doi:10.1140/epja/i2018-12578-2
  [arXiv:1707.03598 [hep-ph]].


\bibitem{Wang:2017azm} 
  W.~Wang, Z.~P.~Xing and J.~Xu,
  Eur.\ Phys.\ J.\ C {\bf 77}, no. 11, 800 (2017)
  doi:10.1140/epjc/s10052-017-5363-y
  [arXiv:1707.06570 [hep-ph]].


\bibitem{Gutsche:2017hux} 
  T.~Gutsche, M.~A.~Ivanov, J.~G.~Körner and V.~E.~Lyubovitskij,
  Phys.\ Rev.\ D {\bf 96}, no. 5, 054013 (2017)
  doi:10.1103/PhysRevD.96.054013
  [arXiv:1708.00703 [hep-ph]].


\bibitem{Li:2017pxa} 
  H.~S.~Li, L.~Meng, Z.~W.~Liu and S.~L.~Zhu,
  Phys.\ Lett.\ B {\bf 777}, 169 (2018)
  doi:10.1016/j.physletb.2017.12.031
  [arXiv:1708.03620 [hep-ph]].


\bibitem{Guo:2017vcf} 
  Z.~H.~Guo,
  Phys.\ Rev.\ D {\bf 96}, no. 7, 074004 (2017)
  doi:10.1103/PhysRevD.96.074004
  [arXiv:1708.04145 [hep-ph]].


\bibitem{Lu:2017meb} 
  Q.~F.~Lü, K.~L.~Wang, L.~Y.~Xiao and X.~H.~Zhong,
  Phys.\ Rev.\ D {\bf 96}, no. 11, 114006 (2017)
  doi:10.1103/PhysRevD.96.114006
  [arXiv:1708.04468 [hep-ph]].


\bibitem{Xiao:2017udy} 
  L.~Y.~Xiao, K.~L.~Wang, Q.~f.~Lu, X.~H.~Zhong and S.~L.~Zhu,
  Phys.\ Rev.\ D {\bf 96}, no. 9, 094005 (2017)
  doi:10.1103/PhysRevD.96.094005
  [arXiv:1708.04384 [hep-ph]].


\bibitem{Sharma:2017txj} 
  N.~Sharma and R.~Dhir,
  Phys.\ Rev.\ D {\bf 96}, no. 11, 113006 (2017)
  doi:10.1103/PhysRevD.96.113006
  [arXiv:1709.08217 [hep-ph]].


\bibitem{Ma:2017nik} 
  Y.~L.~Ma and M.~Harada,
  J.\ Phys.\ G {\bf 45}, no. 7, 075006 (2018)
  doi:10.1088/1361-6471/aac86e
  [arXiv:1709.09746 [hep-ph]].


\bibitem{Yu:2017zst} 
  F.~S.~Yu, H.~Y.~Jiang, R.~H.~Li, C.~D.~Lü, W.~Wang and Z.~X.~Zhao,
  Chin.\ Phys.\ C {\bf 42}, no. 5, 051001 (2018)
  doi:10.1088/1674-1137/42/5/051001
  [arXiv:1703.09086 [hep-ph]].


\bibitem{Meng:2017dni} 
  L.~Meng, H.~S.~Li, Z.~W.~Liu and S.~L.~Zhu,
  Eur.\ Phys.\ J.\ C {\bf 77}, no. 12, 869 (2017)
  doi:10.1140/epjc/s10052-017-5447-8
  [arXiv:1710.08283 [hep-ph]].


\bibitem{Cui:2017udv} 
  E.~L.~Cui, H.~X.~Chen, W.~Chen, X.~Liu and S.~L.~Zhu,
  Phys.\ Rev.\ D {\bf 97}, no. 3, 034018 (2018)
  doi:10.1103/PhysRevD.97.034018
  [arXiv:1712.03615 [hep-ph]].


\bibitem{Shi:2017dto} 
  Y.~J.~Shi, W.~Wang, Y.~Xing and J.~Xu,
  Eur.\ Phys.\ J.\ C {\bf 78}, no. 1, 56 (2018)
  doi:10.1140/epjc/s10052-018-5532-7
  [arXiv:1712.03830 [hep-ph]].


\bibitem{Xiao:2017dly} 
  L.~Y.~Xiao, Q.~F.~Lü and S.~L.~Zhu,
  Phys.\ Rev.\ D {\bf 97}, no. 7, 074005 (2018)
  doi:10.1103/PhysRevD.97.074005
  [arXiv:1712.07295 [hep-ph]].


\bibitem{Yao:2018zze} 
  X.~Yao and B.~Müller,
  Phys.\ Rev.\ D {\bf 97}, no. 7, 074003 (2018)
  doi:10.1103/PhysRevD.97.074003
  [arXiv:1801.02652 [hep-ph]].


\bibitem{Yao:2018ifh} 
  D.~L.~Yao,
  Phys.\ Rev.\ D {\bf 97}, no. 3, 034012 (2018)
  doi:10.1103/PhysRevD.97.034012
  [arXiv:1801.09462 [hep-ph]].


\bibitem{Ozdem:2018uue} 
  U.~Özdem,
  J.\ Phys.\ G {\bf 46}, no. 3, 035003 (2019)
  doi:10.1088/1361-6471/aafffc
  [arXiv:1804.10921 [hep-ph]].


\bibitem{Ali:2018ifm} 
  A.~Ali, A.~Y.~Parkhomenko, Q.~Qin and W.~Wang,
  Phys.\ Lett.\ B {\bf 782}, 412 (2018)
  doi:10.1016/j.physletb.2018.05.055
  [arXiv:1805.02535 [hep-ph]].


\bibitem{Dias:2018qhp} 
  J.~M.~Dias, V.~R.~Debastiani, J.-J.~Xie and E.~Oset,
  Phys.\ Rev.\ D {\bf 98}, no. 9, 094017 (2018)
  doi:10.1103/PhysRevD.98.094017
  [arXiv:1805.03286 [hep-ph]].


\bibitem{Li:2018epz} 
  R.~H.~Li and C.~D.~Lu,
  arXiv:1805.09064 [hep-ph].


\bibitem{Zhao:2018mrg} 
  Z.~X.~Zhao,
  Eur.\ Phys.\ J.\ C {\bf 78}, no. 9, 756 (2018)
  doi:10.1140/epjc/s10052-018-6213-2
  [arXiv:1805.10878 [hep-ph]].


\bibitem{Xing:2018bqt} 
  Y.~Xing and R.~Zhu,
  Phys.\ Rev.\ D {\bf 98}, no. 5, 053005 (2018)
  doi:10.1103/PhysRevD.98.053005
  [arXiv:1806.01659 [hep-ph]].


\bibitem{Zhu:2018epc} 
  R.~Zhu, X.~L.~Han, Y.~Ma and Z.~J.~Xiao,
  Eur.\ Phys.\ J.\ C {\bf 78}, no. 9, 740 (2018)
  doi:10.1140/epjc/s10052-018-6214-1
  [arXiv:1806.06388 [hep-ph]].


\bibitem{Ali:2018xfq} 
  A.~Ali, Q.~Qin and W.~Wang,
  Phys.\ Lett.\ B {\bf 785}, 605 (2018)
  doi:10.1016/j.physletb.2018.09.018
  [arXiv:1806.09288 [hep-ph]].


\bibitem{Liu:2018euh} 
  M.~Z.~Liu, Y.~Xiao and L.~S.~Geng,
  Phys.\ Rev.\ D {\bf 98}, no. 1, 014040 (2018)
  doi:10.1103/PhysRevD.98.014040
  [arXiv:1807.00912 [hep-ph]].


\bibitem{Xing:2018lre} 
  Z.~P.~Xing and Z.~X.~Zhao,
  Phys.\ Rev.\ D {\bf 98}, no. 5, 056002 (2018)
  doi:10.1103/PhysRevD.98.056002
  [arXiv:1807.03101 [hep-ph]].


\bibitem{Bediaga:2018lhg} 
  R.~Aaij {\it et al.} [LHCb Collaboration],
  arXiv:1808.08865.


\bibitem{Wang:2018duy} 
  W.~Wang and R.~Zhu,
  Int.\ J.\ Mod.\ Phys.\ A {\bf 34}, no. 31, 1950195 (2019)
  doi:10.1142/S0217751X19501951
  [arXiv:1808.10830 [hep-ph]].


\bibitem{Dhir:2018twm} 
  R.~Dhir and N.~Sharma,
  Eur.\ Phys.\ J.\ C {\bf 78}, no. 9, 743 (2018).
  doi:10.1140/epjc/s10052-018-6220-3


\bibitem{Berezhnoy:2018bde} 
  A.~V.~Berezhnoy, A.~K.~Likhoded and A.~V.~Luchinsky,
  Phys.\ Rev.\ D {\bf 98}, no. 11, 113004 (2018)
  doi:10.1103/PhysRevD.98.113004
  [arXiv:1809.10058 [hep-ph]].


\bibitem{Jiang:2018oak} 
  L.~J.~Jiang, B.~He and R.~H.~Li,
  Eur.\ Phys.\ J.\ C {\bf 78}, no. 11, 961 (2018)
  doi:10.1140/epjc/s10052-018-6445-1
  [arXiv:1810.00541 [hep-ph]].


\bibitem{Zhang:2018llc} 
  Q.~A.~Zhang,
  Eur.\ Phys.\ J.\ C {\bf 78}, no. 12, 1024 (2018)
  doi:10.1140/epjc/s10052-018-6481-x
  [arXiv:1811.02199 [hep-ph]].


\bibitem{Li:2018bkh} 
  G.~Li, X.~F.~Wang and Y.~Xing,
  Eur.\ Phys.\ J.\ C {\bf 79}, no. 3, 210 (2019)
  doi:10.1140/epjc/s10052-019-6729-0
  [arXiv:1811.03849 [hep-ph]].


\bibitem{Meng:2018zbl} 
  L.~Meng and S.~L.~Zhu,
  Phys.\ Rev.\ D {\bf 100}, no. 1, 014006 (2019)
  doi:10.1103/PhysRevD.100.014006
  [arXiv:1811.07320 [hep-ph]].


\bibitem{Cerri:2018ypt} 
  A.~Cerri {\it et al.},
  CERN Yellow Rep.\ Monogr.\  {\bf 7}, 867 (2019)
  doi:10.23731/CYRM-2019-007.867
  [arXiv:1812.07638 [hep-ph]].


\bibitem{Gutsche:2018msz} 
  T.~Gutsche, M.~A.~Ivanov, J.~G.~Körner, V.~E.~Lyubovitskij and Z.~Tyulemissov,
  Phys.\ Rev.\ D {\bf 99}, no. 5, 056013 (2019)
  doi:10.1103/PhysRevD.99.056013
  [arXiv:1812.09212 [hep-ph]].


\bibitem{Shi:2019fph} 
  Y.~J.~Shi, Y.~Xing and Z.~X.~Zhao,
  Eur.\ Phys.\ J.\ C {\bf 79}, no. 6, 501 (2019)
  doi:10.1140/epjc/s10052-019-7014-y
  [arXiv:1903.03921 [hep-ph]].


\bibitem{Shi:2020qde} 
  Y.~J.~Shi, W.~Wang, Z.~X.~Zhao and U.~G.~Meißner,
  arXiv:2002.02785 [hep-ph].


\bibitem{Hu:2020mxk} 
  X.~H.~Hu, R.~H.~Li and Z.~P.~Xing,
  doi:10.1140/epjc/s10052-020-7851-8
  arXiv:2001.06375 [hep-ph].

\end{thebibliography}
\end{document}